\documentclass{appolb}
\usepackage{graphicx}

\usepackage{amsmath}
\usepackage{amssymb}
\usepackage{graphicx}
\usepackage{dsfont}
\usepackage{dcolumn}
\usepackage{units}
\usepackage{bm}
\usepackage{wasysym}
\usepackage{multirow}
\usepackage{slashed}
\usepackage[usenames]{color}
\usepackage[colorlinks=true,linkcolor=blue,citecolor=blue,urlcolor=blue]{hyperref}
\usepackage{cite}

      \newcommand{\mS}{\mathcal{S}}
      \newcommand{\mD}{\mathcal{D}}
      \newcommand{\mT}{\mathcal{T}}

\begin{document}


\title{Tetraquarks from the Bethe-Salpeter equation
\thanks{Presented by G. Eichmann at the workshop "Excited QCD 2015", Tatranska Lomnica, Slovakia, March 8-14, 2015.}
}

\author{Gernot Eichmann, Christian S. Fischer, Walter Heupel
\address{Institut f\"{u}r Theoretische Physik, Justus-Liebig-Universit\"{a}t Giessen, \\ 35392 Giessen, Germany}}

\maketitle

\begin{abstract}
We present a numerical solution of the four-quark Bethe-Salpeter equation for a scalar tetraquark.
We find that the four-body equation dynamically
generates pseudoscalar poles in the Bethe-Salpeter amplitude. The sensitivity to the pion poles leads to a light isoscalar tetraquark mass $M_\sigma \sim 400$ MeV,
which is comparable to that of the $\sigma/f_0(500)$. The masses of its multiplet partners $\kappa$ and $a_0/f_0$ follow a similar pattern,
thereby providing support for the tetraquark interpretation of the light scalar nonet.
\end{abstract}


\section{Introduction}

  The notion of tetraquarks has undergone a significant paradigm change in recent years.
  There is now increasing evidence for four-quark states in the heavy charm and bottom spectrum, with
  $X(3872)$, $Y(4260)$ and the charged $Z$ states as the prime candidates~\cite{XYZ}.
  Their internal decomposition is under debate: are they `compact' tetraquarks with equally distributed constituents?
  Or do they rather prefer to be in a diquark-antidiquark or a meson-molecule configuration?

  The oldest tetraquark candidates are those in the light scalar meson nonet: the
  $\sigma/f_0(500)$, $\kappa(800)$ and $a_0/f_0(980)$, which
  do not fit into the conventional meson spectrum. Why are the $a_0$ and $f_0$ mass-degenerate, with masses close to $K\bar{K}$ threshold?
  Why are their decay widths so different from $\sigma$ and $\kappa$?
  And why are these states so light? In the quark-model counting, scalar mesons should be $p$ waves with one unit of orbital angular momentum. Hence, they should have
  masses of about $1.2 \dots 1.5$ GeV, similar to their axial-vector and tensor counterparts.

  A possible explanation has been proposed long ago~\cite{tetraquark}: what if these `mesons' were actually tetraquarks in the form of diquark-antidiquark states?
  Two scalar diquarks form again a nonet, but the mass ordering is reversed: the $\sigma$ would be the lightest state made of up and down quarks only,
  whereas $f_0$ and $a_0$ would be heaviest and mass-degenerate because they carry two strange quarks.
  This would also explain the decay widths: $f_0$ and $a_0$ are close to $K\bar{K}$ threshold and therefore narrow, but
  $\sigma$ and $\kappa$ can simply fall apart into two pions or a pion and kaon, respectively, without exchanging gluons.
  In that case the actual $q\bar{q}$ ground states would be indeed the `first radially excited nonet' with masses in the $1.3 \dots 1.5$ GeV region.
  The non-$q\bar{q}$ interpretation of the light scalar nonet is supported by a variety of approaches such as
  unitarized ChPT, quark models, or the extended linear $\sigma$ model~\cite{approaches}.
  The status of tetraquarks at large $N_c$ is discussed in Ref.~\cite{large-nc}; see also Jos\'e Pelaez' contribution to these proceedings.

            \begin{figure}[t!]
            \centerline{%
            \includegraphics[width=12.5cm]{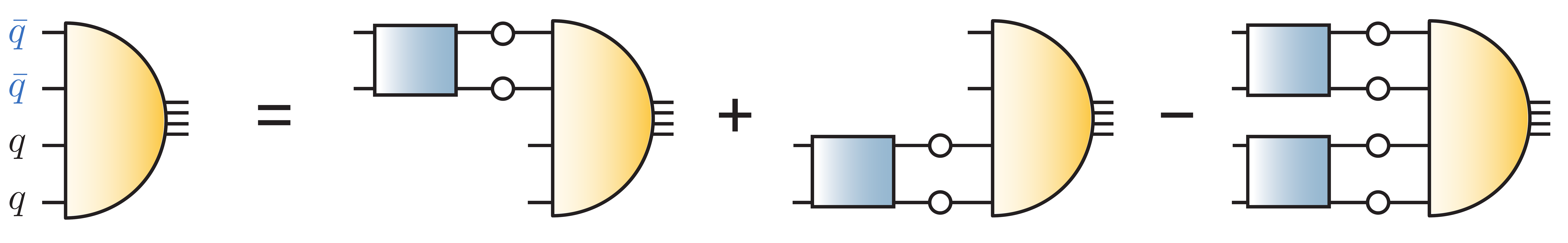}}
            \caption{Four-body BSE for a tetraquark.}
            \label{fig:4b-bse}
            \end{figure}

\section{Bethe-Salpeter approach}

  Irrespective of their internal decomposition,
  what we can say with certainty is the following: if four-quark states exist, they should appear in QCD's spectral representation and produce poles in the eight-quark correlator,
  or more precisely, in the connected  $qq\overline{qq}$ scattering matrix $T$.
  Looking for bound-state poles in higher $n-$point functions is the essence of the Bethe-Salpeter approach:
  at the pole position, the $T-$matrix becomes the product of two Bethe-Salpeter amplitudes, and from the onshell propagator pole we can read off the mass of the state.

  The $T-$matrix can be determined from the scattering equation or inhomogeneous Bethe-Salpeter equation (BSE). It has the form $T=K+K\,G_0\,T$, where $G_0$ is the product of four dressed quark propagators
  and $K$ is the four-quark interaction kernel. The scattering equation at the pole reduces to the homogeneous BSE shown in Fig.~\ref{fig:4b-bse},
  which has a solution only if the $T-$matrix has a pole.
  The kernel is the sum of two-, three- and four-body irreducible interactions (we already omitted the three- and four-body pieces in Fig.~\ref{fig:4b-bse}),
  and the specific form of the pair interactions is necessary to prevent overcounting~\cite{Khvedelidze:1991qb,Heupel:2012ua}.
  The equation in the figure is written in the `diquark-antidiquark' topology (12)(34); there are two further permutations (23)(14), (31)(24) with meson-meson topologies.

  The two-body kernel that appears here must be consistent with the underlying quark-gluon structure to preserve QCD's chiral symmetry.
  A feasible way to achieve this is to employ a rainbow-ladder kernel, where the two-quark interaction is simplified to an iterated gluon exchange,
  and solve the Dyson-Schwinger equation (DSE) for the quark propagator with this input.
  As reviewed elsewhere~\cite{RL}, the phenomenology of pseudoscalar and vector mesons, as well as baryon octet and decuplet ground states,
  can be reasonably well described in such a setup.
  This implies not only mass spectra but also their form factors and other properties,
  and it extends to charmonium and bottomonium spectra.
  Calculations beyond rainbow-ladder are underway and aim, for example, to include the `pion-cloud' corrections to form factors~\cite{pion-cloud}.

  On the other hand, there are many observables where gluon exchange alone does \textit{not} provide the answer, for instance:
  scalar and axialvector mesons (the `$p$ waves' in the quark model), heavy-light mesons, or excited hadrons.
  When viewed as a $q\bar{q}$ state, rainbow-ladder produces a $\sigma-$meson mass of $600 \dots 700$ MeV. Its status beyond rainbow-ladder is presently unclear:
  in some recent calculations the mass goes down even further~\cite{BRL-sigma}, which would again provide support for a $q\bar{q}$ interpretation.
  On the other hand, these studies do not only produce a light isoscalar but also a mass-degenerate light isotriplet,
  which leads back to the `wrong' mass ordering: $\{\sigma, a_0\} \rightarrow \kappa \rightarrow f_0$.
  Hence, there are most likely further effects at play.

\section{Tetraquarks: two-body equation}

     As a first step towards solving the four-body problem, let us take a simpler route that is motivated by the quark-diquark model for baryons~\cite{QDQ}.
     When three-body interactions are neglected, the three-body BSE analogous to Fig.~\ref{fig:4b-bse}
     can be simplified to a quark-diquark BSE where gluons no longer appear. Instead, quarks interact with scalar and axialvector diquarks via quark exchange.
     It turns out that the quark-diquark approach, when employing rainbow-ladder DSE and BSE solutions for its ingredients,
     provides a rather good approximation for the three-body equation regarding nucleon and $\Delta$ properties~\cite{QDQ-vs-3b}.

\begin{figure}[t!]
            \centerline{%
            \includegraphics[width=12.5cm]{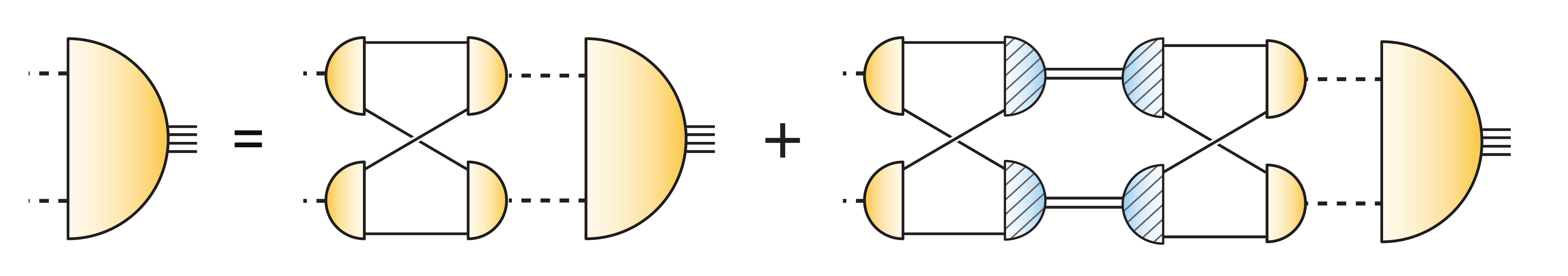}}
            \caption{Two-body meson-meson/diquark-antidiquark BSE. Single, double and dashed lines are quark, diquark and meson propagators, respectively.
                      The blobs are the respective Bethe-Salpeter amplitudes.
                     All ingredients are dressed.}
            \label{fig:2b-bse}
            \end{figure}

     Based on this observation, one can proceed in an analogous way with the four-body equation. Fig.~\ref{fig:4b-bse}
     can be rewritten as a Faddeev-Yakubovsky equation~\cite{faddeev-yakubowski}, and after assuming separability for the $qq$ and $q\bar{q}$ scattering matrices it reduces
     to a coupled diquark-antidiquark/meson-meson equation where the interaction takes place via quark exchange~\cite{Heupel:2012ua}. In fact, the equations can be combined into a single
     meson-meson equation (shown in Fig.~\ref{fig:2b-bse}) but not vice versa. This already hints at the meson-molecule nature of the system, where diquarks only appear as internal admixtures.
     Using rainbow-ladder DSE and BSE solutions for the ingredients, the solution for the $\sigma$ mass was found to be $\sim 400$ MeV~\cite{Heupel:2012ua}.

     The evolution of the isoscalar-scalar mass as a function of the input current-quark mass is shown in the left panel of Fig.~\ref{fig:masses}.
     It is noteworthy that the omission of the diquark diagram in Fig.~\ref{fig:2b-bse} changes almost nothing, so
     the diquark admixture is practically irrelevant.
     The light $\sigma$ mass is therefore tied to the behavior of the light pions which are QCD's Goldstone bosons in the chiral limit:
     the tetraquark mass lies just below the pseudoscalar-meson threshold, except for very low quark masses where it becomes a resonance.
     Moreover, the calculation also predicts a relatively light $ss\bar{s}\bar{s}$ state as well as an all-charm tetraquark below $2m_{\eta_c}$.

\section{Tetraquarks: four-body equation}

     How reliable are these results given the approximations that were employed?
     To find out, let us go back to the original four-body equation.
     A scalar $q\bar{q}$ state is predominantly in $p$ wave, but a scalar four-quark state forms an $s-$wave orbital ground state.
     Employing a rainbow-ladder kernel is therefore well motivated from the aforementioned meson and baryon studies, and its reliability can be judged from their results.

     The main difficulty in solving the system in Fig.~\ref{fig:4b-bse} is the complicated structure of the tetraquark amplitude.
     It is a five-point function and depends on four independent momenta.
     The amplitude can be decomposed into two color structures and 256 Dirac-Lorentz tensors, and the resulting 512 Lorentz-invariant dressing functions depend on nine independent variables.
     It is practically impossible to solve this equation on present computers, so we must resort to approximations.
     Since they should respect the symmetries of the system, we proceed as follows:

            \begin{figure}[t!]
            \centerline{%
            \includegraphics[width=12.5cm]{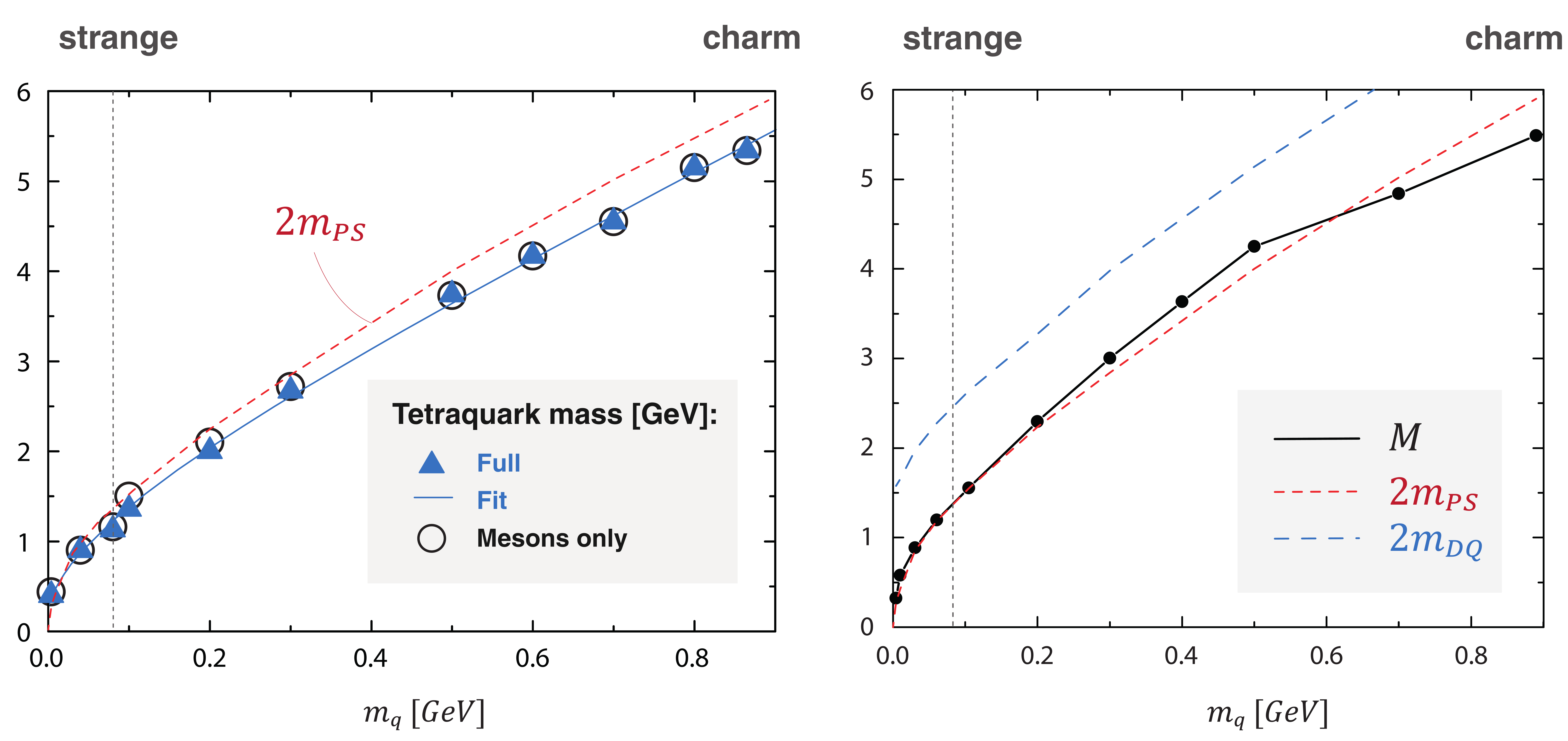}}
            \caption{Current-quark mass dependence of the scalar-isoscalar tetraquark mass from the two-body BSE~\cite{Heupel:2012ua} (\textit{left panel})
                     and from the four-body BSE (\textit{right panel}). The dashed lines are the respective thresholds.}
            \label{fig:masses}
            \end{figure}

     (i) We keep the 16 ($s$-wave) Dirac-Lorentz tensors that do not depend on any relative momentum.
           This is a closed set under Fierz transformations, so it contains `diquark-antidiquark'-like structures such as $C^T \gamma_5 \otimes \gamma_5 C$ and $C^T \gamma^\mu \otimes \gamma^\mu C$
           in the (12)(34) topology, but via Fierz identities it also generates meson-meson-like structures $\sim \gamma_5\otimes\gamma_5$ in the crossed channels.

     (ii) We take into account both color structures, which can be written in either of the Fierz-equivalent topologies:
           diquark-antidiquark ($3\otimes\bar{3}$, $6\otimes\bar{6}$) or meson-meson ($1\otimes 1$, $8\otimes 8$).

\begin{figure}[t!]
            \centerline{%
            \includegraphics[width=12.5cm]{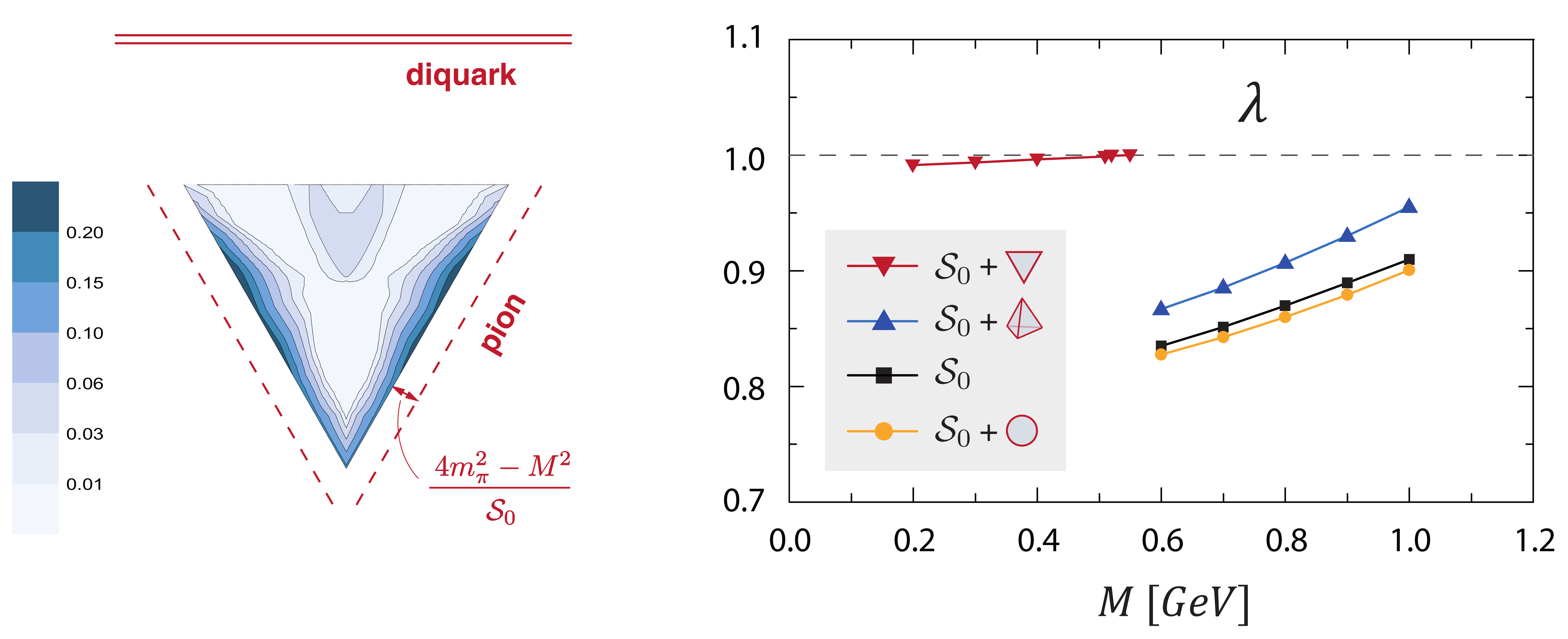}}
            \caption{\textit{Left panel:} Mandelstam triangle whose axes (not shown) are the doublet variables in Eq.~\eqref{multiplets}.
                     The contour plot refers to the magnitude of the leading dressing function.
                     \textit{Right panel:} BSE eigenvalue after switching on different multiplets.}
            \label{fig:ev}
            \end{figure}

     (iii) We arrange the nine momentum variables into irreducible multiplets of the permutation group $S_4$~\cite{s4}, which
           allows us to switch on groups of variables separately in the solution process and thereby judge their importance.
           The nine variables can be grouped into a symmetric singlet $\mathcal{S}_0$, a doublet $\mD_0$, and two triplets $\mT_1$, $\mT_2$:
           \begin{equation}\label{multiplets}
               \mS_0 = \frac{1}{4}\,(p^2+q^2+k^2)\,, \quad \mD = \frac{1}{4\mS_0}\left[\begin{array}{c} \sqrt{3}\,(q^2-p^2) \\ p^2+q^2-2k^2 \end{array}\right], \quad \dots
           \end{equation}
           where $p^2$, $q^2$, $k^2$ are the Mandelstam variables for the three topologies.
           Hence, the dressing functions can be written as $f_i=f_i(\mS_0, \mD, \mT_1, \mT_2)$.
           The doublet phase space that remains invariant under the equation forms the interior of a triangle bound by the lines $p^2=0$, $q^2=0$ and $k^2=0$ (see left panel in Fig.~\ref{fig:ev}),
           whereas the triplets form a tetrahedron and a sphere.

     The resulting eigenvalue of the four-body BSE is plotted in Fig.~\ref{fig:ev}. Since the equation has the form $K\Psi=\lambda\Psi$,
     one can extract the mass of the state by tracking the largest eigenvalue of the kernel as a function of the total momentum squared: $\lambda(P^2=-M^2)=1$.
     If we allow the system to generate a momentum dependence in $\mS_0$ only, the eigenvalue (second curve from below) would cross the line $\lambda=1$ at $M \sim 1400$ MeV.
     This is essentially `four times the constituent-quark mass' and would describe a rather simple tetraquark.
     Switching on the triplets does not change the result appreciably, but adding the doublet $\mD$ has a drastic effect:
     the eigenvalue curve is now almost flat and crosses at $M \sim 400 \dots 500$ MeV.

     What happened?
     The exterior of the triangle exhibits
     pion and diquark poles at timelike values of the Mandelstam variables, illustrated by the dashed and double lines in Fig.~\ref{fig:ev}.
     We emphasize that the four-body equation knows a priori \textit{nothing} about pions and diquarks
     but rather generates them dynamically. They influence the behavior of the dressing functions in the interior, which is clearly visible in the contour plot:
     the pion mass is small and the pion poles are close to the triangle, whereas the scalar diquarks have no visible effect because their mass scale $\sim 800$  MeV is much larger.
     If the pion poles are removed by hand via phase space cuts, the uppermost eigenvalue curve in the right panel
     approaches again its companions.

     Therefore, we can confirm our findings from the two-body equation: the system is dominated by pions rather than diquarks.
     While this would be difficult to judge from the tensor structures alone (because they mix under Fierz transformations),
     the momentum dependence of the amplitudes clearly provides such information.
     The equation also produces a physical threshold: if $M \geq 2m_\pi$, the poles enter the integration domain and
     the tetraquark becomes a resonance.
     In that case we must resort to extrapolations to estimate the real part of the mass.
     The current-mass dependence of the scalar-isoscalar tetraquark mass is shown in the right panel
     of Fig.~\ref{fig:masses} and mirrors the result from the two-body equation.
     At the physical $u/d$ mass, our results for $\sigma$, $\kappa$ and $a_0/f_0$ are
     \begin{equation}
         M_\sigma \sim 380\,\,\text{MeV}, \qquad
         M_\kappa \sim 700\,\,\text{MeV}, \qquad
         M_{a_0/f_0} \sim 920\,\,\text{MeV}\,.
     \end{equation}

\section{Summary}

     We have solved the four-quark Bethe-Salpeter equation for a scalar tetraquark.
     The outcome provides a simple dynamical explanation for the light $\sigma$ mass: the system is dominated by pion poles,
     which are dynamically generated in the solution process and drive the tetraquark mass from `four times the constituent-quark mass' to $M_\sigma\sim 400$ MeV.
     The masses for the multiplet partners $\kappa$ and $a_0/f_0$ follow an analogous pattern.
     This provides support for the interpretation of the light
     scalar nonet as tetraquarks. However, they are not diquark-antidiquark states but predominantly `meson molecules',
     which explains their mass ordering as an indirect consequence of QCD's spontaneous chiral symmetry breaking.

    \bigskip

       \footnotesize
       \textbf{Acknowledgments.}
         This work was supported by the German Science Foundation DFG under project number DFG TR-16, by the Helmholtz International Center for FAIR
         within the LOEWE program of the State of Hesse, and by the Helmholtzzentrum GSI.

\end{document}